\documentclass[twocolumn]{aastex631}
\usepackage{algorithm}
\usepackage{algpseudocode}

\newcommand{\jh}[1]{\textbf{\textcolor{blue}{#1}}}
\shorttitle{Disentangling gas components using NMF.}
\shortauthors{De Mijolla et al.}
\graphicspath{{./}{figures/}}
\begin{document}

\title{Disentangling Multiple Emitting Components in Molecular Observations with Non-negative Matrix Factorization}

\correspondingauthor{Serena Viti}
\email{viti@strw.leidenuniv.nl}

\author[0000-0001-8757-4936]{Damien de Mijolla}
\affiliation{Department of Physics and Astronomy, University College London, Gower Street, London WC1E 6BT}

\author[0000-0003-4025-1552]{Jonathan Holdship}
\affiliation{Leiden Observatory, Leiden University, PO Box 9513, NL-2300 RA Leiden, The Netherlands}
\affiliation{Department of Physics and Astronomy, University College London, Gower Street, London WC1E 6BT}

\author[0000-0001-8504-8844]{Serena Viti}
\affiliation{Leiden Observatory, Leiden University, PO Box 9513, NL-2300 RA Leiden, The Netherlands}
\affiliation{Department of Physics and Astronomy, University College London, Gower Street, London WC1E 6BT}
\author[ 0000-0003-0567-8796 ]{Johannes Heyl}
\affiliation{Department of Physics and Astronomy, University College London, Gower Street, London WC1E 6BT}

\begin{abstract}

Molecular emission from the galactic and extragalactic interstellar medium (ISM) is often used to determine the physical conditions of the dense gas. However, even from spatially resolved regions, the observed molecules are not necessarily arising from a single component. Disentangling multiple gas components is often a degenerate problem in radiative transfer studies. In this paper we investigate the use of the non-negative matrix factorization (NMF) approach as a means to recover gas components from a set of blended line intensity maps of molecular transitions which may trace different physical conditions. We run a series of experiments on synthetic datasets designed to replicate conditions in two very different environments: galactic pre-stellar cores and the ISM in high redshift galaxies. We find that the NMF algorithm  often recovers the multiple components resembling those used in the data-generating process, provided that the different components have similar column densities. When NMF fails to recover all the individual components it does however group  together the most similarly emitting ones. We further found that initialisation and regularisation are key factors in the efficiency of the NMF algorithm.

\end{abstract}

\keywords{Molecular gas (1073), Interstellar phases (850), Submillimeter astronomy (1647), Astrostatistics techniques (1886), Astrochemistry (75)}


\section{Introduction}

Understanding the chemical and physical processes regulating the interstellar medium (ISM) is necessary for a thorough understanding of many astronomical processes, such as star formation \citep{VanDishoeck1998} and galaxy quenching \citep{molecules_galaxy_quenching}. As molecular lines are the primary means through which the conditions of the interstellar medium can be probed, their interpretation is of vital importance for accomplishing these goals.

Radiative-transfer modelling is the framework through which radiation emitted by the ISM can be connected back to the physical conditions of the gas it is emitted from. Through computational radiative-transfer modelling codes, such as RADEX \citep{VanderTak2007}, the strength of molecular transitions can be used to constrain the chemistry, temperature, density and other astrophysical conditions of the emitting interstellar medium. This, in turn, then enables a better understanding of the local environment in which the ISM is embedded.

%The interstellar medium (ISM) exhibits a rich chemistry. Depending on the local environment in which it is embedded, the interstellar medium can exist in many different components. These can be dense or diffuse, hot or cold, atomic or molecular and depending on physical conditions will emit differently across molecular and atomic transitions. Through radiative transfer models, the radiation emitted by the interstellar medium can be connected back to the physical conditions of the gas it is emitted from and used to constrain it's chemistry, temperature, density and other astrophysical conditions. This enables molecular lines measurements to act as probes into the physical conditions of the interstellar medium and be used as tools for understanding many astronomical environments such as for example star-formation and galaxy quenching.

Unfortunately, the radiative-transfer modelling problem needing solving when analysing molecular lines is typically degenerate. Because of the large number of free parameters in radiative-transfer models and their similar influence on intensities, usually multiple set of input parameters are capable of reproducing any given set of molecular line intensities. In \cite{Tunnard2016}, it was found that radiative transfer modeling was typically incapable of recovering parameters to better than half a dex. Although such degeneracies can be partially addressed through using rigorous statistical methodologies \citep{Tunnard2015,James_2021} or through incorporating astrochemical knowledge \citep{Viti2017,de_Mijolla_2019}, the interpretation of molecular lines still remains difficult.

%However, the use of molecular lines as gas tracers is often complicated by degeneracies in their analysis. The large number of free parameters within radiative transfer models make it such that often multiple input parameter combinations are capable of reproducing observations. In \cite{Tunnard2016}, it was found that radiative transfer modeling was typically incapable of recovering parameters to better than half a dex. Although such degeneracies can be partially addressed through using rigorous statistical methodologies \cite{Tunnard2015,James_2021} or through incorporating astrochemical knowledge \cite{Viti2017,de_Mijolla_2019}, the analysis of molecular lines still remains particularly challenging.

Notably, the analysis of emission arising from blended components is particularly challenging. In sub-mm astronomy, the spatial resolution limits of modern telescopes mean that we typically do not observe flux from a single gas component but rather from multiple non-resolved components, each with unique physical conditions.  This is particularly the case for extra-galactic observations where objects as large as giant molecular clouds, or even whole sections of a galaxy, will be contained in a single beam. When this is the case, interpreting the physical conditions of the ISM becomes especially difficult as it is unclear what fraction of the measured emission originates from each component or even how many components are responsible for the observed emission. Moreover, as observers increasingly turn their attention towards the analysis of external galaxies, where multiple blended components are expected, mitigating this blending is likely to become an ever more pressing concern. 

%Interpreting such non-resolved molecular line intensities is difficult. The radiative-transfer problem that needs to be solved when inferring the physical conditions of the interstellar medium froma limited number of molecular lines is typically degenerate \cite{Tunnard2016} and these degeneracies becomes much more severe when fitting multiple non-resolved components.

To address the difficulties associated with the  radiative-transfer problem, there is value in approaching the task of disentangling gas components within observations from a more data-driven perspective. In this study, we investigate the use of data-driven approaches for separating the flux emitted by molecular lines into the individual gas components within observations. More precisely, we investigate the use of NMF \citep{NNMFPaper} which is a matrix-factorization algorithm used to decompose a non-negative matrix into a product of non-negative matrices. In NMF, components are estimated globally from all pixels available. As there will typically be far fewer components than pixels, NMF can then exploit the redundancy across pixels to constrain the possible shapes of the components.

Our work is not the first attempt at using data-driven approaches to understand the ISM. For example, there has been a long line of research into using Principal Component Analysis to interpret observations of the ISM \citep{PCAGiantMolecularCloud,PcaOrion,SpitzerPCA}. Perhaps more similar to our work, there have also been efforts into using NMF to interpret interstellar spectra. For example, NMF has been used to analyze interstellar bands \citep{BerneInfraredBands,FoschinoNMF}. Efforts at using NMF in astronomy date as far back as 1997, where positive matrix factorization, a precursor to NMF, was used to interpret molecular observations \citep{NMF1996}. 

%\ddm{However, all of these works as far as we are aware, attempt to use spectral information (redshift) from a single or small number of molecular lines rather than using large quantities of molecular lines as tracers of different components. These past approaches thus primarily use the spectral shape of individual lines to recognize commoving gas components.} 

Specifically, in this study we attempt to answer the question of whether NMF can be used to recover gas components from a set of blended line intensity maps of transitions tracing different physical conditions. Here, unlike previous approaches such as those presented in \cite{NNMFHyperspectra,NMF1996}, our interest is less about identifying co-moving parcels of gas within observations and more about recovering components tracing a common physical environment even if they may not have originated from the same spatial location. Whereas previous approaches have operated on the spectral datacube where frequency information is available and only made use of a small number of line-transitions, here we operate on the derived line-intensity maps for a larger number of lines and in doing so ignore the spectral shape of line-transitions. This means that  our factorization ignore the radial velocity of spectral lines and only use the molecular composition when constructing components. Our hope is then for NMF components to group together all emission arising from gas under similar physical conditions leading to NMF components capturing the archetypical signatures of physical environments in observations.

\section{Data model}\label{secNMF:data_model}

We begin by detailing a model for the blending of components which relates line intensities and spatial locations of components in a given region to the intensities measured in line-intensity maps.

Let us represent a set of molecular line intensity-maps, observed over a region of $N_x$ $\times$ $N_y$ pixels, by a non-negative matrix \jh{$X$} of shape $N_M \times N_p$ whose entries contain the intensities for $N_M$ molecular line transitions measured at the $N_p=N_x \times N_y$ pixels. We assume that multiple components contribute additively to the intensities such that the intensity at every pixel is the sum of the intensities of the separate components. We represent the strength of emission of the components in the various line-transitions by a (non-negative) matrix $W_{true}$ of shape $N_M \times N_C$ whose $N_C$ columns each represent the intensities emitted in the $N_M$ molecular transitions for a component.\par
Similarly, we encode the spatial contributions of each of the $N_C$ components to the $N_p$ pixels into a (non-negative) matrix $V_{true}$ of shape $N_C \times N_{p}$. Spatial contributions are factors between 0 and 1 indicating how much a component contributes to a pixel. The intensity of a given line at a given pixel will simply be the sum of the emission in that line for all components multiplied by their contribution to that same pixel. That is to say the line-intensity maps can be written as
\begin{equation}
    X=W_{true}V_{true}+\Sigma
\label{eq:model}
\end{equation}
where $\Sigma$ is an additional noise term chosen to be Gaussian.

So far, we have neglected the impact of convolution by a telescope beam factor on observations. Provided that all line intensities are downsampled to a common spatial resolution, convolution can be modeled through a matrix multiplication by a convolution matrix K of shape $N_{p} \times N_{p}$. The data-generating process then becomes $X=W_{true}V_{true}K+\Sigma$ which can be written using the associativity property of matrix multiplication as $X=W_{true}V_{conv}+\Sigma$ where $V_{conv}=V_{true}K$. Hence, including convolution by a beam-filling factor does not change the original model  
beyond the replacement of $V_{true}$ with $V_{conv}$.

%Although it may at first seem that the data-generating process described above neglects the effect of convolution by the beam-filling factor on observations. Provided that all line intensity maps are downsampled to a common spatial resolution the convolving by the beam filling factor can be incorporated directly within $V_{true}$. Indeed, provided the effect of convolution is the same on all line-intensity maps, the convolution can be accomplished through a matrix multiplication by a convolution matrix K of shape $N_{p} \times N_{p}$.  By virtue of matrix multiplication being associative the convolution can thus be first applied to the contributions in which case $V_{true}$ are the components convolved by the \ddm{Idea directly modify the data-generating paragraph to incorporate this}.

A number of conditions must be met for our model of line-intensity maps to be accurate. Firstly, molecular transitions must not be optically thick so as for intensities to be (near) additive. Secondly, all components must be purely emitting rather than absorbing as we assume all contributions are positive. Finally, as explained in the previous paragraph, transitions must be preliminarily downsampled to a common spatial resolution for the convolution to be parameterizable by a matrix $K$.

\section{Non-negative matrix factorization}\label{secNMF:nmf}

In this study, we consider the problem of recovering a set of blended components $W_{true}$ and their convolved spatial locations $V_{conv}$ from intensity maps \jh{$X$}, under the assumption that $X=W_{true}V_{conv}+\Sigma$. Although many algorithms could be relevant for this task, here we focus on investigating the non-negative matrix factorization algorithm (NMF) \citep{NNMFOriginal}. Given a matrix \jh{$X$}, the NMF algorithm aims to find non-negative matrices $W$ and $V$ such that $X \approx WV$, where $W$ and $V$ are matrices of shape $N_M \times N_{NMF}$ and $N_{NMF} \times N_p$ and $N_{NMF}$ is the number of components used. In the context of molecular observations, $N_{NMF}$ is the number of physically distinct emitting components assumed to exist within telescope beam. It is a user-defined parameter for the algorithm which should ideally be equal to $N_C$ the true number of components so as for the dimensions of $W$ and $V$ to match those of $W_{true}$ and $V_{true}$. NMF falls within the realm of blind-source separation algorithms which is to say that it is an algorithm making minimal assumptions on the contents of matrices $W$ and $V$.

Formally, the NMF algorithm proceeds by determining non-negative \jh{$W$} and \jh{$V$} minimizing a loss function
\begin{equation}
    L= \|X-W V\|
\end{equation}
where $\|A\|$ is a matrix-norm which is often chosen to be the Frobenius norm $\|A\|_{\text {F}}^{2}=\sum_{i, j} A_{i j}^{2}$. Minimizing the loss function is a non-convex optimization problem which is typically approached using iterative solvers. For this paper, we use the scikit-learn implementation of NMF making use of a coordinate-descent solver \citep{NMFCoordinateDescent} with \jh{$W$} and \jh{$V$} initialized as random matrices.

Matrix factorization is an ill-posed problem. Indeed, if $X=WV$, then it would also be the case that $X=WPP^{-1}V$ for any invertible matrix P (of shape $N_{C} \times N_{C}$) and so if $W$ and $V$ are solutions to the matrix factorization problem $WP$ and $P^{-1}V$ will also be solutions, provided they are non-negative matrices. Because of this, there is no formal guarantees that the NMF algorithm applied to line-intensity maps \jh{$X$} will converge towards $W_{true}$ and $V_{conv}$, even in the limit of infinite data.

In NMF, to further constrain the matrix factorization problem towards desirable solutions, it is common to add to the loss function additional regularization terms on $W$ and $V$. In such cases, the loss function minimized by NMF becomes:

\begin{equation}
    L= \|X-W V\|_{F}+\alpha_{1}\|W\|_{R}+\alpha_{2}\|V\|_{R}
\end{equation}

In addition to regularization, the initialization of matrices $W$ and $V$ can impact the convergence of NMF and hence the matrices retrieved by the algorithm. In this project, we initialize the algorithm with non-negative random matrices as defined in the scikit-learn NMF documentation.

It is also worth mentioning that the NMF algorithm, like many other matrix factorization algorithms, is most effective when there are significantly fewer components than molecular lines; that is to say when $N_{C}<<N_{M}$. This is because NMF relies on finding a set of components capturing the most information about observations X which requires grouping together correlated molecular lines. In the limit where the number of components matches the number of molecular lines observed ($N_{C}=N_{M}$), the NMF loss is trivially minimized by setting \jh{$W$} to the identity matrix (i.e. one component reconstruct each line) and \jh{$V$} to be equal to \jh{$X$}. Such a solution perfectly reconstructs observations but brings no useful information about the correlations amongst molecular lines.

%We focus on data. We treat this task as a a blind source separation problem. That is to say we make minimal assumptions about $W$ and $V$ when retrieving them.

\section{Synthetic data generation}\label{secNMF:generation}
In order to assess the use of NMF in interpreting molecular line-intensity maps, we test its effectiveness at retrieving components from synthetic line-intensity maps. Synthetic data allow us to have exact knowledge of the ground-truth components and therefore evaluate how closely they are recovered by NMF. Our aim when constructing these synthetic observations is to approximately reproduce components and observations found within real-world datasets. Our aim is not to build extremely accurate simulations. As such, we have striven for simplicity over complexity in our data-generating process.

We construct synthetic line-intensity maps of transitions of CO, CS, HCN and HCO$^+$ (see Section 5), using numpy \citep{numpy}, by first using chemical and radiative transfer models to produce emission profiles ($W_{true}$) from chosen physical conditions. We then specify an arbitrary spatial location of the emission of each component ($V_{true}$). Finally, we select a noise level ($\Sigma$) and convolution due to the beam-filling factor ($K$) to produce the maps ($X$) following Eq~\ref{eq:model} as discussed in Section \ref{secNMF:data_model}. 

The matrix $W_{true}$ contains the emission of each component for the molecular lines considered in $X$. For our experiments, we select a temperature and density for each component and the intensity of molecular lines are then determined using the radiative transfer modelling code RADEX \citep{VanderTak2007}. RADEX requires a gas density and temperature as well as the column density of the emitting species. Since the emission is proportional to the column density, and we choose the noise level, the absolute value of the column density is completely arbitrary and has no effect on the experiments. However, the differences in column density between components is observationally important so we use an astrochemical model to obtain reasonable values.

To do this, we use the astrochemical code UCLCHEM \citep{Holdship2017} to find the steady state abundances of our species for the chosen temperature and density of each component, assuming standard Milky Way values for the cosmic-ray ionization rate and UV field of, respectively, $1.3\times 10^{-17} \rm s^{-1}$ and 1 Draine (equivalent to 2.74$\times$10$^{-3}$ erg/s/cm$^2$). We convert abundances to molecular column densities by finding the factor that would convert the component with the highest CO abundance to a CO column density of $10^{19} \rm cm^{-2}$. We then scale the other components by the same factor so that their column density ratios are equal to their abundance ratios. Therefore, whilst our absolute column density values are arbitrary, their ratios are informed by chemistry.

The matrix $V_{true}$ encodes how the emission of each component varies spatially from the peak value given by RADEX. We model the spatial contribution of a component at every pixel as a 2d Gaussian whose dimensions correspond to the x and y coordinates of the pixel. That is to say, the contribution at a pixel $v_{xy}$ is $v_{xy}= A\times N(\mu,\sigma)$ evaluated at the x,y coordinates of the pixel center. In this data-generating process, the mean $\mu$ of a component controls its location, the covariance $\sigma$ its extent and the amplitude $A$ its relative strength. For all components, the amplitude A is set such that $v_{xy}=1$ at the peak location of the component. The entries of $V_{true}$ are then the contribution $v_{xy}$ of the $N_{p}$ pixels for the $N_D$ components.

%It is worthwhile acknowledging that our parametrization of $V_{true}$ is somewhat restrictive. Indeed, it can only produce elliptical emission components and so does not incorporate the flexibility required for reproducing complex observables such as for example a galaxy's spiral arms. However, within the context of NMF, such a simplistic parametrization has it's merits. As the NMF algorithm treats all pixels independently, any geometrical information between pixels is ignored by the algorithm and as consequence, the geometrical realism of the synthetic datasets has very limited impact on results. \ddm{Instead, it is the level of spatial overlap between components which is important to reproduce in our synthetic observations.}\ddm{Scrambling the pixels location within the image would not affect results.}

The telescope beam-response is approximated by a Gaussian kernel K. After creating noiseless line-intensity maps according to $W_{true}V_{true}K$, in our final step gaussian noise N is added to line-intensity maps such that every map has a signal-to-noise-ratio of 100 with gaussian noise independently, identically, and uniformly added to all pixels.

\section{Experiments}\label{secNMF:experiments}

 We now present our experiments on applying NMF to synthetic observations (with results plotted using matplotlib \citep{matplotlib}). All experiments are carried out using a common blueprint for the characteristics/properties of the line-intensity map. We choose to produce synthetic observations of two environments: protostellar cores and high redshift galaxies. We arbitrarily pick the molecular transitions from the observational study of the nearby galaxy NGC 1068 from \cite{Viti2014}.  While the choice of this particular sample of lines was arbitrary, we note that it contains some of the most observed molecules in both galactic as well as extragalactic environments. All line-intensity maps are constructed on a 50 $\times$ 50 dimensional grid (i.e. $50\times50$ pixels), for the transitions observed in \cite{Viti2014}, to which Gaussian noise is added such that maps have a mean per-pixel SNR of 100 and to which a convolution with a Gaussian kernel of width 1/10 of the overall width of maps has been applied. Although this blueprint is only one amongst many possible parametrizations, we found the conclusions of our study to be robust to changes to the blueprint. 

%We now present our experiments on applying NMF to synthetic observations. For these experiments, we have constructed line-intensity maps on a 50 $\times$ 50 dimensional grid, for the transitions observed in \cite{Viti2014}, to which Gaussian noise is added such that maps have a mean per-pixel SNR of 100 and to which a convolution with a Gaussian kernel of width 1/10 of the overall width of maps has been applied.\ddm{The specific parameters for constructing the line-intensity maps were not found to significantly affect results and so line-intensity maps could have been constructed differently without affecting conclusions.}\ddm{The choice of hyperparmaeters in the construction of these line-intensity maps }

In the following experiments, before application of NMF, we always apply a series of preprocessing steps to the line intensity maps. These preprocessing steps would likely need to be reproduced when working on real observations.  As a first step, in order to satisfy the non-negativity requirement of NMF, all negative-entries in the line-intensity maps are zeroed-out. As a second step, so as for the NMF loss to equally weight all line-intensity maps, we rescale line-intensity maps prior to running NMF through multiplication by a scaling factor such that, after multiplication, all line-intensity maps have an equal average flux of unity. After the NMF algorithm has finished running, this scaling factor is erased from the NMF components through dividing entries in $W$ by the scaling factor associated to their transition. 

As described in Section~\ref{secNMF:nmf}, there are two regularisation terms that can be used to control how the algorithm converges. To investigate the sensitivity of our approach to this regularisation, we try 3 different regularisation approaches (regularise $W$ on its own, $V$ on its own, or both $W$ and $V$ simultaneously) and for each of these regularisation approaches we try 21 different values of $\alpha$.

Whilst testing the sensitivity to regularisation is important, there is a further confounding factor. The result of NMF varies depending on the initialisation of matrices $W$ and $V$. To capture this variation, we repeat each regularisation set up 300 times with different random matrix intialisations which is achieved by passing init=``random'' in the scikit-learn NMF implementation. This allows us to be sure that where one regularisation set up performs better than another, it is not due to random fluctuations because we can test whether it is consistent across 300 repeats.

Overall, each experiment is performed with 3 different regularisation approaches at 21 different regularisation strengths repeated 300 times each for a total of 18,900 runs. To analyse this population of NMF runs, we have created a goodness-of-fit metric condensing all information about each individual run into a single number quantifying the level of agreement between recovered and true components. This metric is defined as the mean-absolute error (MAE) between the true components emission profiles $W_{true}$ and those  retrieved by the NMF run ($W$). However, since components returned by NMF can only be correct up to a scaling factor, we rescale the emission profiles to have an average flux (averaged over all transitions) of unity before calculating the MAE. This definition of MAE means that the line intensities ($W$) of components retrieved by a run with a MAE of 0.1 will on average differ from the ground-truth intensities ($W_{true}$) by $10\%$ of the average line-intensity of their closest matching ground-truth component. Finally, since any of the three NMF components could correspond to any of the three true components, we choose the permutation that gives the lowest MAE. We show in Appendix \ref{appendix:pseudocode} the pseudo-code summarising the overall procedure for generating results as described in this section.

%We found that, because of the under-constrained nature of the NMF factorization problem, the convergence of the NMF algorithm was sensitive to the regularization employed and the initialization of matrices W and V. Rerunning the algorithm with a modified initialization or regularization could yield, after convergence, dramatically different matrices W and V. As such, since each NMF run could only probe a unique initialization and regularization, individual runs were incapable on their own of fully characterizing the distribution of matrices expected from running NMF. 

%As such, so as for experiments to give a fair and complete insight into the performance and stability of NMF, we characterize the performance of NMF in the aggregate by running a population of runs in which hyperparameters (regularization and initialization) are varied. By doing this, we are capable not only of understanding how well NMF fits observations but also what type of prior information regularization terms and matrice initializations actually encode. Moreover, as no ground-truth components are available for real data, we are interested in understanding how to select regularization and initial conditions without access to ground-truth data and what impact sub-optimal choices may have. Further experiments on real datasets, where ground-truth data does not exist, may then find it beneficial to use a set of hyperparameters which was found to work well on synthetic data closely resembling the dataset being studied.

\subsection{Protostellar environments}\label{secNMF:low-overlap}
\begin{figure}
\includegraphics[width=\linewidth ,page=7,trim={5cm 0.5cm 13cm 4cm},clip]{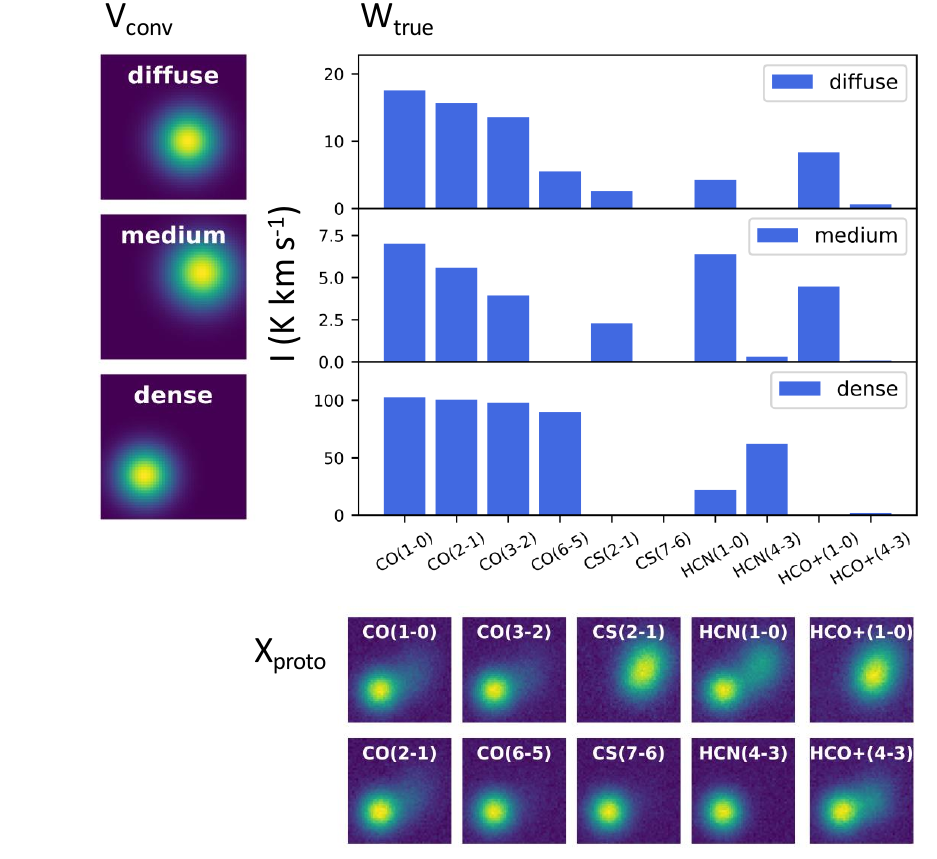}
\caption{Synthetic line-intensity maps of molecular clouds ($X_{proto}$ : bottom) alongside the emission profile ($W_{true}$ : top-right) and convolved (unitless) spatial contribution ($V_{conv}$ : top-left) of each components contributing to the maps.}
\label{fig:clouds_data}
\end{figure}

In this section we present insights derived from applying NMF to the first of two sets of mock line-intensity maps $X_{proto}$, which models emission from blended protostellar environments. The maps in $X_{proto}$ are generated from three components, each meant to approximate a gas component found in star forming regions. Generated components are i) An extended ``low density'' component (T=20 K and n=$10^4 \rm cm^{-3}$), ii) a ``medium density'' component (T=10 K and n=$10^5 \rm cm^{-3}$), and iii) a ``dense'' warmer component (T=100 K and n=$10^7 \rm cm^{-3}$). These three idealized components may, for example, be representing the surrounding ISM, the envelope, and the more central compact region of a protostellar core.

The synthetic-line intensity maps ($X_{proto}$) as well as emission profiles and convolved contributions of emitting components are shown in Figure \ref{fig:clouds_data}. From this Figure, we see some moderate spatial overlap between components as well as some components emitting more strongly than others. Both of these factors make it difficult, at least visually, to recover components from the line-intensity maps motivating the use of specialized retrieval algorithms.

\begin{figure*}
\includegraphics[width=\linewidth ,page=9,trim={4.4cm 0cm 5cm 0cm},clip]{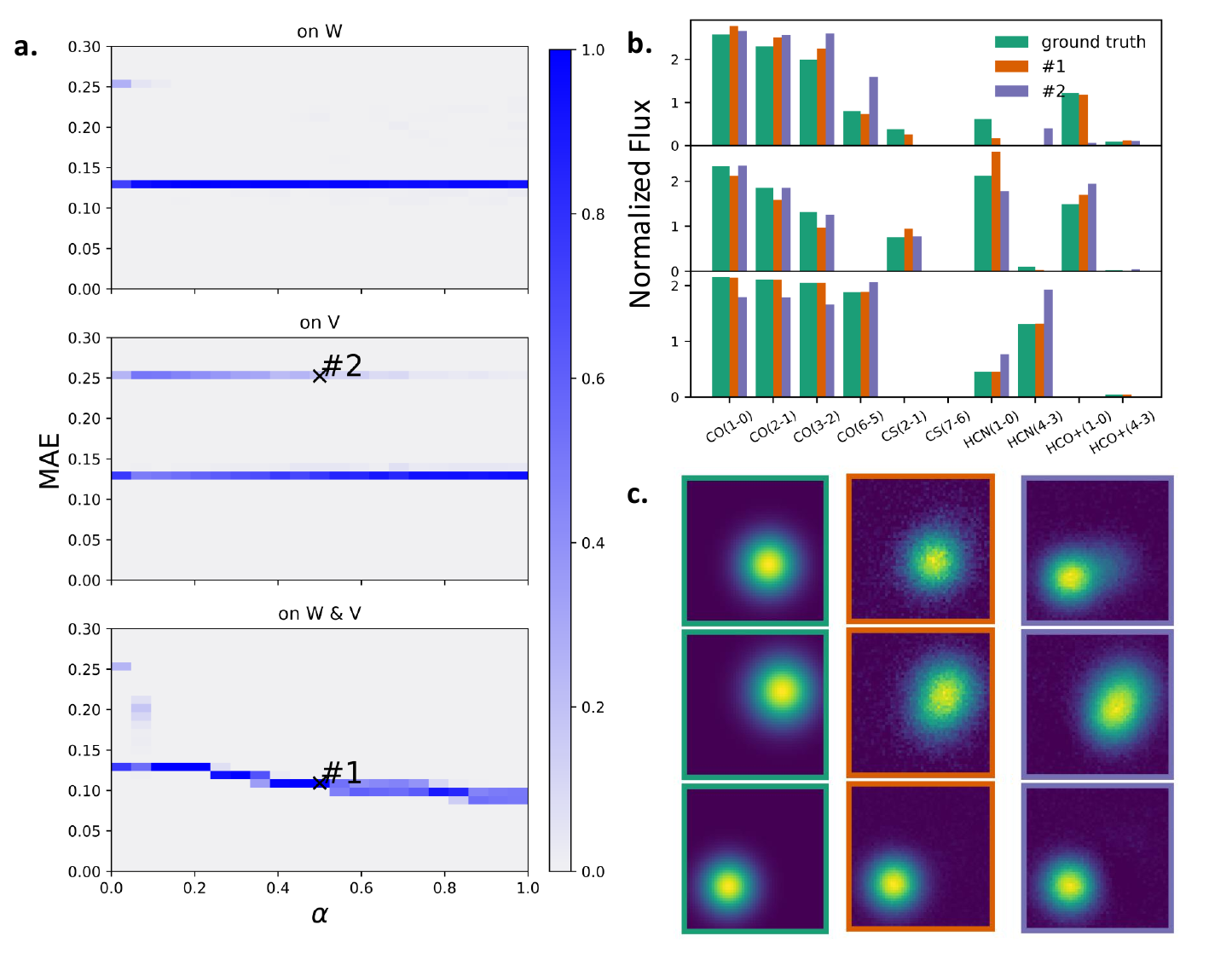}
\caption{\textbf{a)} Binned density plots characterizing the performance of a population of three-component NMF runs on $X_{proto}$ in which x-axis coordinates quantify the amplitude of the regularization term and y-axis coordinates the goodness of fit of the retrieved components as defined by the MAE metric introduced in Section \ref{secNMF:experiments}. Each panel considers a different type of regularization: i) only on $W$ (top panel), ii) only on $V$ (central panel), or iii) on both $W$ and $V$ (bottom panel). \textbf{b)} Emission profiles of the true components alongside those of two runs in the population whose location in the scatter-plot are represented by markers \#1 \& \#2. \textbf{c)} Convolved spatial contributions of the same two runs.}
\label{fig:clouds_combined}
\end{figure*}

We show in Figure \ref{fig:clouds_combined} outputs obtained from running a population of NMF runs on $X_{proto}$. Runs in the population use three NMF components, the same number as the number of chemical components used in the modelling, which means that the algorithm should have enough capacity to fit all components.
Figure \ref{fig:clouds_combined}a (on the left) characterises the goodness of fit of NMF across this population of runs. Each panel of the figure represents a binned density plot of runs in which  the x-axis coordinates quantify the amplitude of the regularisation term and the y-axis coordinate the goodness of fit of the retrieved components as defined by the MAE metric introduced in Section \ref{secNMF:experiments}. The binned density plot captures the probability distribution of the goodness-of-fit metric at a given regularisation value (i.e. at a fixed x-axis value). Each panel can be viewed as a succession of vertically displayed histograms capturing the probability distribution in the MAE metric for a regularisation level specified by the axis. Here, the probability distributions capture the variability which arises due to the random initialisation for the initial guesses of matrices $W$ and $V$. Each panel considers a different type of regularisation: i) only on $W$ (top panel), ii) only on $V$ (central panel), or iii) on both $W$ and $V$ (bottom panel).
%The top-left figure shows a scatter plot characterizing the goodness of fit across this population of runs in which every point represents an NMF run whose x-axis coordinate quantifies the amplitude of the regularization term, y-axis coordinate the goodness of fit of the retrieved components as defined by the MAE metric introduced in Section \ref{secNMF:experiments}, and color the type of regularization used.
Figure \ref{fig:clouds_combined} b and c focus-in on two runs within the population and compare the relative line strengths ($W$) and spatial contributions ($V$) of these runs to those of the ground-truth components. The locations of these runs within the density plots are shown by markers \#1 and \#2. These two runs are selected for further study from the population of NMF runs because they represent two extremes (in terms of agreement with $W_{true}$) amongst all runs.

We find an excellent agreement between the components from run \#1, one of the better fitting runs in the population, and the true components. Although there are minor mismatches between true and recovered components, such as most of the HCN(1-0) originating from the diffuse (top-panel) true component being mistakenly attributed to the middle-panel retrieved component, the recovered components are by and large in excellent agreement with the true gas components. However other runs in the population differ more strongly from the true components. Run \#2 is an extreme example of such a run with one of the lowest y-values across the population. This run exhibits a much higher level of disagreement with the true components, with in particular its top-panel component not matching  well  any of the true components. The disagreement observed stems from the run learning a different decomposition into components than that used to generate the data. In this decomposition, the top panel component captures emission from both the top and bottom panel of the true components. Such a decomposition, while different from the data-generating decomposition, was found to match the lower MAE runs (e.g. run\#1) in terms of its quality of fit to the line-intensity maps. The decomposition found in run\#2 is thus an equally valid decomposition that cannot be ruled-out from studying the quality of the match between $X_{proto}$ and the factorized approximation $WV$.

By comparing the distribution of the y-coordinates of runs in the models population to the y-values of runs \#1 and \#2, we can understand how effective and reliable NMF is at recovering components closely matching with the underlying component. We observe that the vast majority of runs, irrespective of regularisation, have y-values comparable to \#1, indicating that most runs retrieve components matching well with the underlying components. While poor-match runs comparable to \#2 do exist, these runs are relatively rare. This seems to suggest that, at least for the considered dataset, NMF is an effective, albeit imperfect, algorithm for recovering the underlying components.

We can also study the impact of hyperparameters on NMF results. Regularization appears to have a beneficial effect on the agreement between recovered and ground-truth components with even small amounts of regularization on $W$ being sufficient to steer the factorization away from the higher MAE solutions. Whilst regularization on $V$ does not have as strong of an impact, it also improves the match between true and recovered gas components as can be seen from how the best performance is obtained with high-levels of regularization on both $W$ and $V$. For now we do not discuss how the choice of  the number of NMF components affects results (but such a discussion can be found in Section \ref{highz_two_component}). However, it is worth acknowledging that we expect performance to degrade when using a sub-optimal number of components.

%From figure \ref{fig:clouds_scatter}, we see that the goodness of fit is affected by the regularization employed. Regularization appears to be beneficial as higher regularization models have better agreement between retrieved and true components. Although the exact form of regularization does affect the quality of the convergence, in general, all forms of regularization are helpful as seen from how the performance is at it's worst when no regularization is employed (x=0). It is also clear that the random initialization of matrices can play a huge role into the convergence of the algorithm. This can be seen from the large scatter amongst models of the same type at the same regularization. At least for this dataset, it seems that xx amount of regularization is perhaps best.

% \begin{figure*}
% \centering
% \includegraphics{figures/chap3/molecular_clouds/zoom_phases.pdf}
% \caption{TBD}
% \label{fig:clouds_zoom_phases}
% \end{figure*}

\subsection{High-z galaxy (three-component fits)}\label{secNMF:high-overlap}
Disentangling emission arising from blended
components is of course a problem that exacerbates the further away the object is.  In this  section, we therefore study the performance of the NMF algorithm on a synthetic dataset approximating line-intensity maps from a high-redshift galaxy $X_{gal}$. Unlike in the previous experiment, where components had only moderate spatial overlap, in this scenario all components peak in the same location -  the center of the galaxy - but have different spatial extents. This high-overlap between components complicates the retrieval process as there are fewer pixels which can clearly be attributed back to a unique component. This dataset thus constitutes a more challenging test for NMF but an important one as the NMF algorithm will only be practical if it can reliably produce useful retrievals across a range of different scenarios.

The components used in this run are a less dense component (T=10 K and n=$10^4 \rm cm^{-3}$), a medium density moderately extended and warm component (T=30 K and n=$10^5 \rm cm^{-3}$), and a dense compact hotter central component (T=50 K and n=$10^7 \rm cm^{-3}$). We note that, for a spatially unresolved high redshift galaxy, our three gas components are definitely  more arbitrary in nature than in our previous example. Nevertheless, we may identify the first component with the average cold molecular gas, typical of a Giant Molecular Cloud, the second component with dense molecular gas heated by either nearby star formation or an AGN, and the third component with compact very dense gas found in the proximity of the nucleus of a galaxy.

\begin{figure}
\includegraphics[width=\linewidth ,page=8,trim={5cm 0.5cm 13cm 4cm},clip]{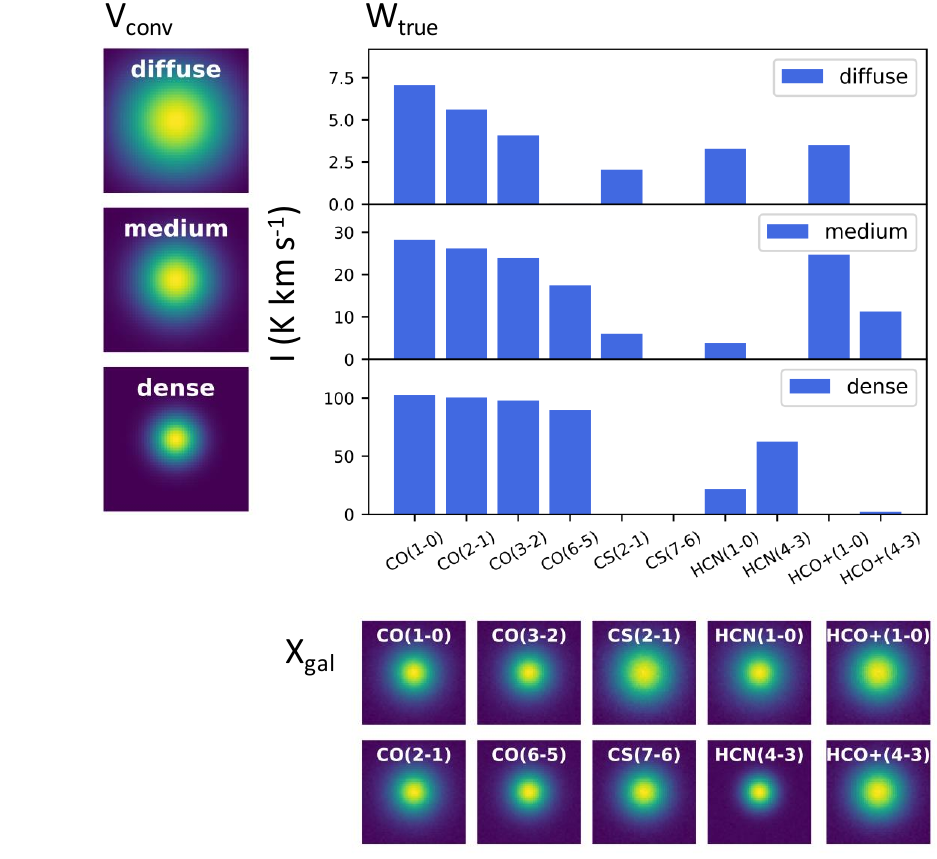}

\caption{Synthetic line-intensity maps of high-redshift galaxy ($X_{gal}$ : bottom) alongside the emission profile ($W_{true}$ : top-right) and convolved (unitless) spatial contribution ($V_{conv}$ : top-left) of each components contributing to the maps.}
\label{fig:high_z_data}
\end{figure}

\begin{figure*}
\includegraphics[width=\linewidth ,page=10,trim={4.4cm 0cm 5cm 0cm},clip]{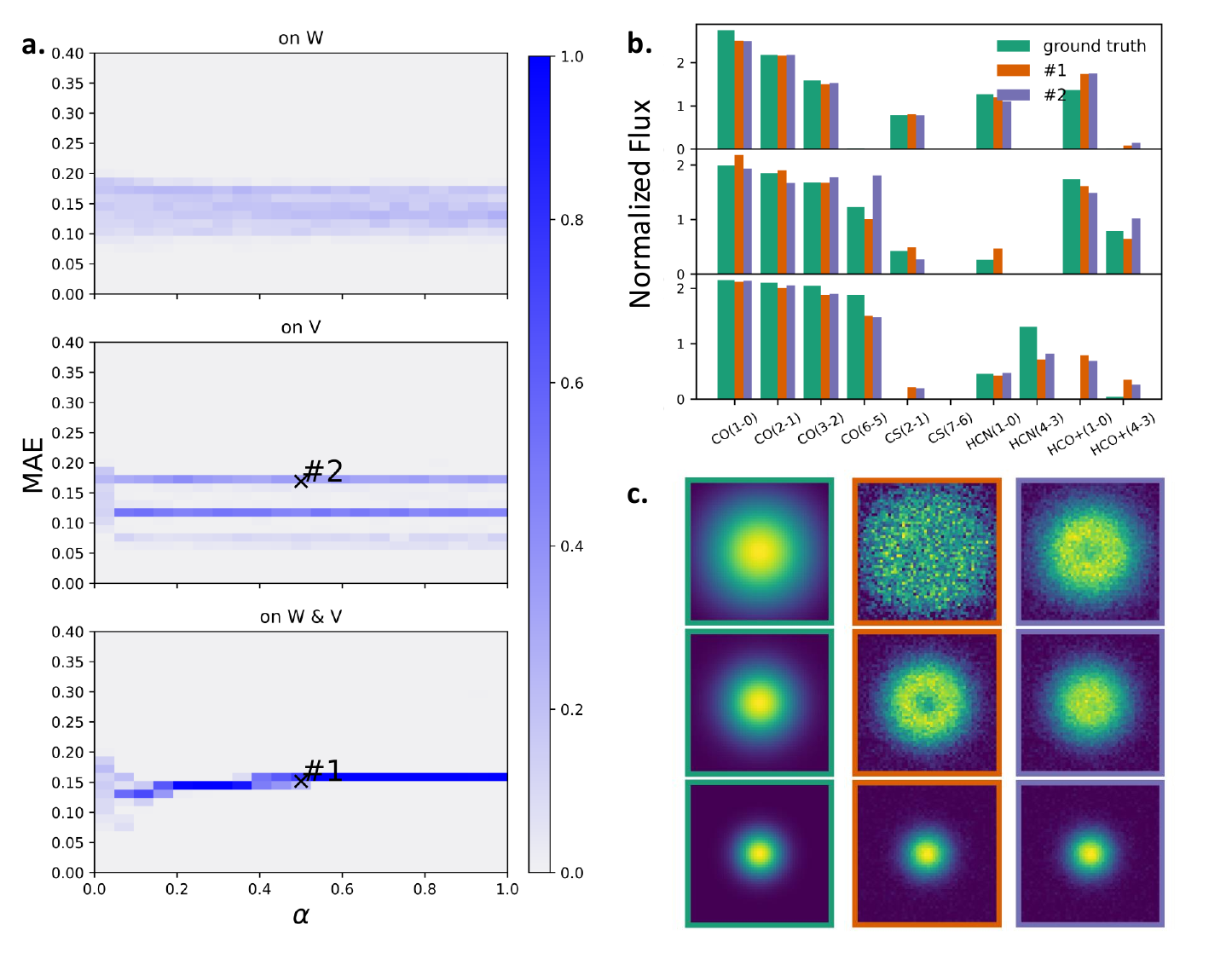}
\caption{\textbf{a)} Binned density plots characterizing the performance of a population of three-component NMF runs on $X_{gal}$ in which x-axis coordinates quantify the amplitude of the regularization term and y-axis coordinates the goodness of fit of the retrieved components as defined by the MAE metric introduced in Section \ref{secNMF:experiments}. Each panel considers a different type of regularization: i) only on $W$ (top panel), ii) only on $V$ (central panel), or iii) on both $W$ and $V$ (bottom panel). \textbf{b)} Emission profiles of the true components alongside those of two runs in the population whose location in the scatter-plot are represented by markers \#1 \& \#2. \textbf{c)} Convolved spatial contributions of the same two runs.}
\label{fig:high_z_combined}
\end{figure*}

Once again, we validate the algorithm through plotting i) line-intensity maps and associated components, ii) density plots showing the MAEs for a population of three-component runs, and iii) the retrieved components for two runs whose locations in the scatter plot are shown by markers \#1 and \#2. These plots are shown in Figure \ref{fig:high_z_data} and \ref{fig:high_z_combined}. NMF runs use three NMF components to fit the line-intensity maps which is a number equal to the number of components within observations and as many as was used in the previous experiment.

We can see from these figures that there is a relatively good level of agreement between the true components and those recovered in runs \#1 and \#2, with runs having at most a MAE of 0.2 (which represents an average of 20\% difference in intensity between the true line intensities and those produced by the NMF algorithm). This suggests that for this high-overlap scenario, NMF can approximately recover the underlying components. However, the mismatch between true and recovered components appears larger here (at least in terms of contributions) than in the previous low-overlap scenario (Section \ref{secNMF:low-overlap}). In particular, the algorithm seems to struggle with decomposing the fully overlapping central component, favouring a decomposition into spatially non-overlapping components, despite the high overlap betweenthe underlying components. This can be seen by how, in both runs, emission arising from more extended components in the central portion of the observations is misattributed to the less extended retrieved component (bottom panel). This in turn then leads to a mismatch between the true and retrieved emission profile for this less extended components as well as the creation of donut-like holes in the spatial profiles of the other more extended components (at the location of misattributed emission). 

Overall, these results highlight the presence of degeneracies when doing retrievals of observations containing highly overlapping components. When such high-spatial overlap components exist, the NMF algorithm struggles to recover the proper overlap between components. Furthermore, no choice of regularisation appears capable of lifting these degeneracies. Nevertheless, recovered components still match fairly well with observations.

\subsection{High-z galaxy (two-component fits)}\label{highz_two_component}

In real-world applications, the number of components used in the NMF retrieval will need to be estimated from observations. However, because of  differences in physical conditions and chemical make-up, different spatial locations will emit in different molecular transitions  leading to the existence of a higher number of components. In such cases, to maintain the regime of fewer NMF components than observed line-transitions, NMF retrievals must be run with fewer NMF components than required to fully capture observations. The aim is then not to retrieve the true components but instead to retrieve useful components. We examine here how NMF behaves in this regime.

\begin{figure*}
\includegraphics[width=\linewidth ,page=6,trim={3.0cm 8cm 5cm 0cm},clip]{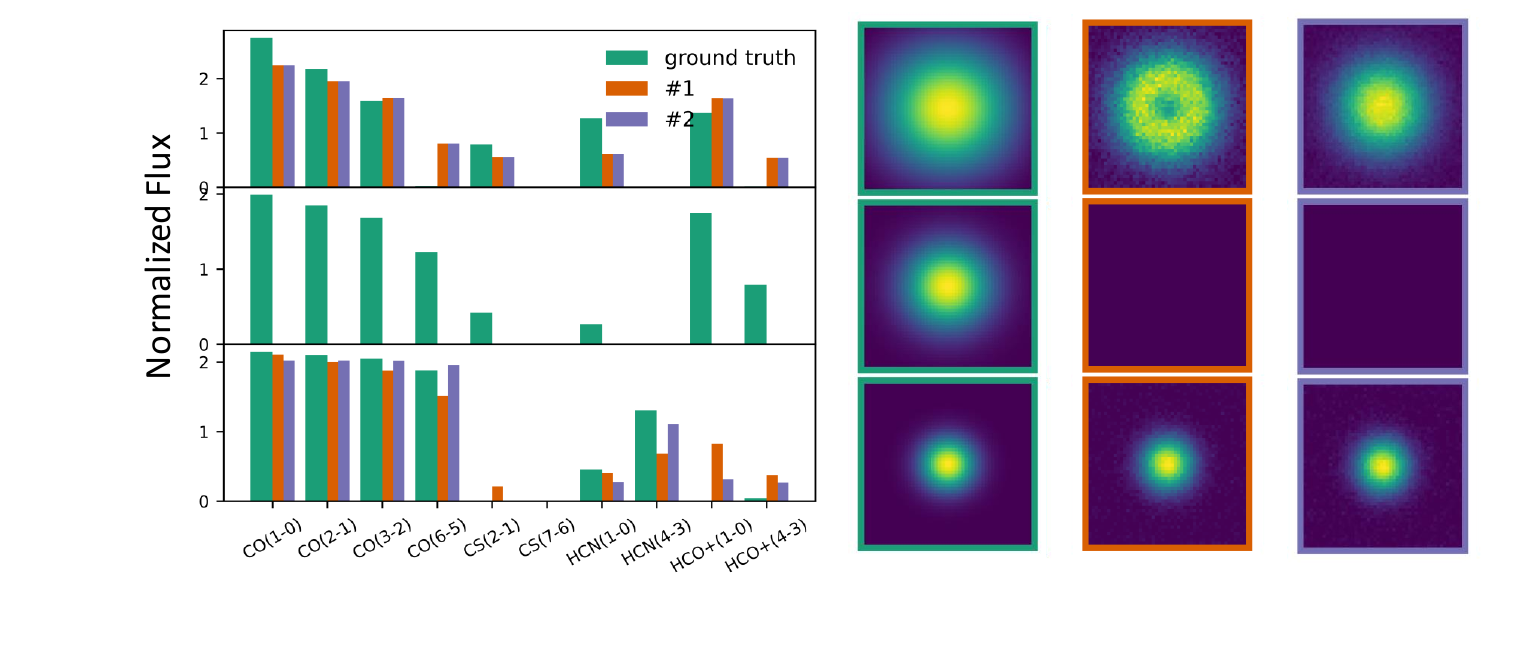}
\caption{Emission profiles and (unitless) convolved spatial contributions of the true components alongside those of two random runs in a population of two-component runs fitted to $X_{gal}$.}
\label{fig:high_z_2}
\end{figure*}

To do this, we run NMF on our high-z line intensity maps but using only two components instead of the three components actually required for reproducing the line-intensity maps. This provides a test for the behaviour of the NMF algorithm when data is more complex than the available number of components. Figure \ref{fig:high_z_2} shows the output of such two-component runs. Here, we do not show density plots since our goodness of fit metric is not applicable when the number of retrieved and ground truth gas components differ.

Comparing these runs to their three-component counterparts (Figure \ref{fig:high_z_combined}) reveals that the two-component runs seem to only differ from the three-component runs in their treatment of the top and middle-panel components. Because the two-component runs are incapable of fully capturing the emission from all three components, they group together the top and middle-panel components and instead return a component whose emission profile is approximately a mixture of the emission from these two components. On the other hand, the lowest-panel component of the two-component runs, characterizing the densest component, is almost identical to the analogous component in the three-component runs. In fact, similarly to the three-component runs, it (mistakenly) captures emission from other gas-phases within the central region leading to a similar mismatch between retrieved and true dense component and a similar (incorrect) donut shape spatial emission in other components.

Conceptually, it appears that because the two-component NMF runs did not have enough capacity to reproduce all three components, they selectively grouped the two most similar components - the top and middle panel components - into a single ``archetypical component'' capturing the average emission across both phases whilst dedicating the remaining component to the densest component because it emitted more uniquely. Such behaviour suggests that, when lacking enough components to fit observations, the NMF algorithm will fall-back on grouping together the most similar components into archetypical components. This means that the algorithm could allow for synthesizing highly-complex observations into a manageable number of components for human interpretation.

\subsection{High-z galaxy - dependence on the number of components}

In the previous section we studied the fit to our high-z environment $X_{gal}$, retrieved by a two-component NMF model. We found that the two-component model did a good job at reproducing line-intensity maps despite these technically containing three components. In this  section we explicitly study the effect of varying the number of components $N_C$ used in the NMF algorithm on the quality of fit to line-intensity maps. It is worth noting that whereas previous sections have evaluated the NMF algorithm by assessing the quality of its retrieved components, here we evaluate NMF by assessing how well it reproduces the line-intensity maps (irrespective of whether this emission arises from components resembling the true gas phases). This better matches the reality of fitting an unknown number of components where the only data we have are the line-intensity maps.

\begin{figure}
\includegraphics[width=\linewidth]{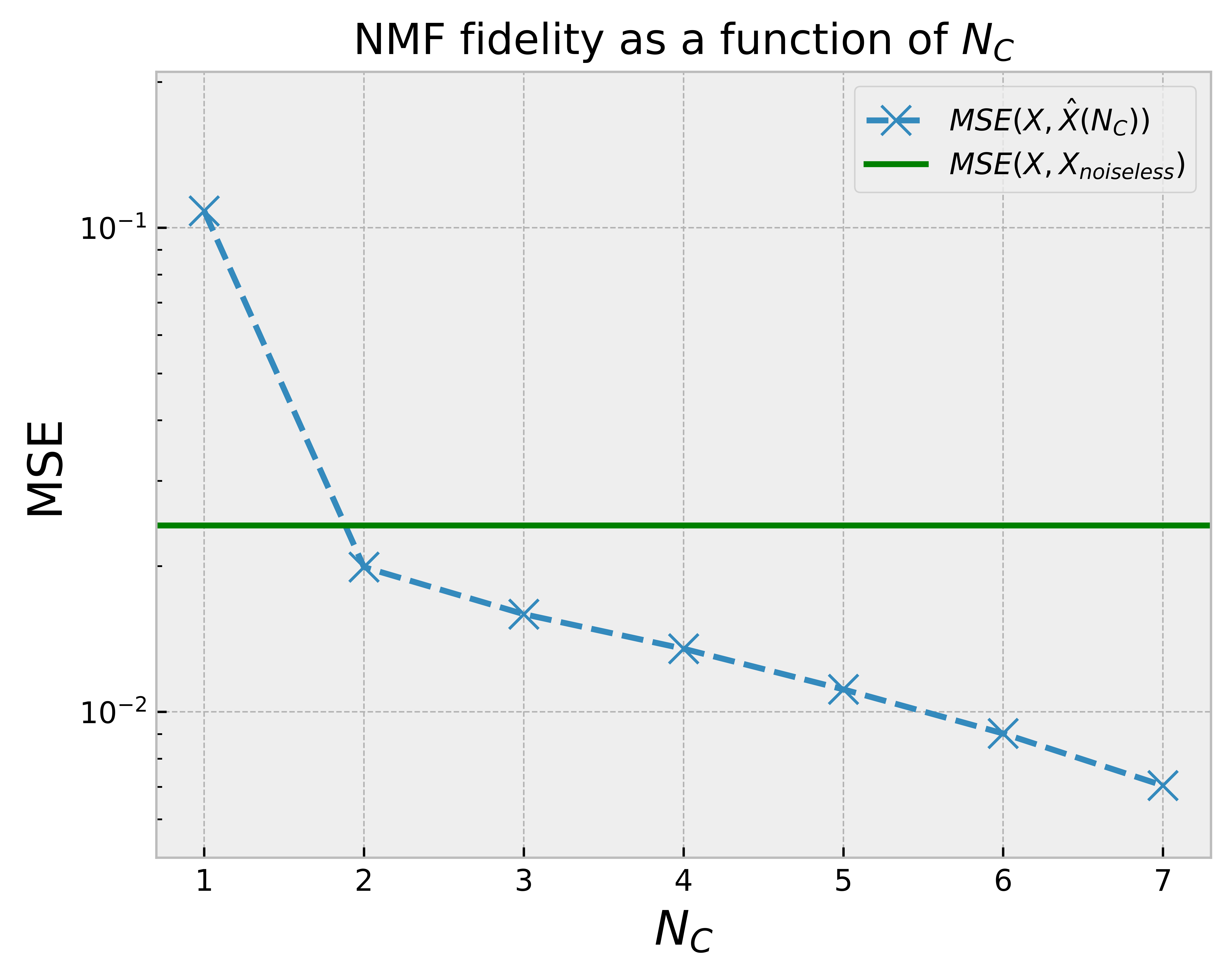}
\caption{Mean-squared error (MSE) between the NMF fits to $X_{gal}$ (obtained as $WV$ and denoted as $\hat{X}(N_C)$) and $X_{gal}$ for varying number of NMF components $N_C$ (blue) where MSE is plotted on log-scale. As a baseline we also plot the MSE between $X_{gal}$ and the noiseless version of $X_{gal}$ (green).}
\label{fig:mse_fit}
\end{figure}

To show this, in Figure \ref{fig:mse_fit} we plot the mean-squared error (MSE) between the NMF fits to $X_{gal}$ (obtained as $WV$) and $X_{gal}$ for different values of $N_C$, the number of components in the NMF (shown on the x-axis). Here, to calculate the MSE, we first normalise each line-intensity map in order to prevent any single line-intensity map from dominating over others in the calculation of the MSE, then flatten the line-intensity maps into a 1D vector and finally calculate $MSE = \frac{1}{N} \sum_{i=1}^{N}(x_i-\hat{x}_i)^2$. We also show, as a baseline, the MSE between $X_{gal}$ and the noiseless version of $X_{gal}$ which represents some ideal MSE value for a model reproducing the ground-truth gas phases and none of the noise.

Interestingly, we find that a two-component NMF fit, similar to that explored in the previous subsection, does a good job at matching the emission in observations $X_{gal}$ despite using fewer components than the number of gas phases present in the observations. This is particularly interesting because it may be conceptually similar to analysing an emission map as a set of archetypical components (gas originating from a starburst ring; dense gas surrounding an AGN; etc) as it is commonly done, rather than considering its true physical gas components. In fact, we find that the two-component fit has a lower MSE than the baseline representing the MSE solely due to noise. This is due to the NMF decomposition overfitting the noise in the line-intensity maps. From this figure, we further find that increasing the number of components in the NMF algorithm leads to further overfitting to the noise in observations, as evidenced by the seemingly linear decrease in MSE with increasing dimensionality of $N_C$.

Since the noise level can be estimated from the map itself, this provides a method for estimating the number of components one should use to analyse observations. A similar plot can be produced for a real line-intensity map and the optimal number of components is the number that causes the MSE of the NMF algorithm to reach the true noise level. Crucially, this would be the number of gas components that are sufficiently distinct for the algorithm to separate, and not the true number of physical components. However, this would be useful since, even when analysing archetypical components, there is often a question of how many to include. This approach should be further tested by analysing low resolution maps of objects where high resolution imaging has uncovered distinct components.

\subsection{Discussion and conclusions}\label{secNMF:discussion}

We have run a series of experiments aiming at recovering the underlying components within observations of molecular line-intensity maps.
These experiments were run on a set of synthetic datasets designed to replicate conditions expected from real observations while also capturing some diversities of environments where the algorithm could be applicable.
Because NMF ignores the spatial locations of pixels and would perform equally well if pixels were scrambled, we did not put a strong emphasises on reproducing the geometrical structure of real molecular maps and instead focused on reproducing the expected overlap between components.
To capture the diversity of environments we built two synthetic line-intensity maps on which to test our observations: one with a moderate overlap between components and a second dataset with a high overlap.
We validated the effectiveness of NMF on these synthetic environments through studying the algorithm's ability to recover the components used in the data-generating process.
Because the convergence of the NMF algorithm was found to be sensitive to hyperparameters, we have characterised the performance across a range of hyperparameters rather than for a single set of hyperparameters.

The results of our investigations were as follows.
We found that the NMF algorithm  often recover components resembling those used in the data-generating process. We found that when the NMF algorithm lacked the capacity to reproduce all components, it would group together the most similarly emitting components into one archetypical component.
However, in general, the matrix-factorization problem solved by NMF did not have a unique solution and so its convergence to a set of components closely resembling the underlying components was not guaranteed.
Instead, initialisation and regularisation played a crucial role towards influencing the type of solution the algorithm converged towards, with sometimes very different solutions depending on the hyperparameters.
This sensitivity of NMF to model hyperparameters is obviously not ideal as it means that the algorithm can sometimes recover components that match poorly with observations.
In general we also found that the NMF algorithm struggled with recovering the proper spatial overlap of overlapping components.

Our experiments provide a proof-of-concept towards the applicability of NMF for analysing line-intensity maps.
However, some differences do still exist between real data and the data within our experiments.
One such difference is that our mock line-intensity maps shared a common beam size. When working with real data, it will be crucial to down-sample observations so as to guarantee that parcels of emitting gas have the same spatial extent across all line transition maps.
Another difference is that because of effects such as optical depth and differences in local physical conditions, it is unlikely that real data will be perfectly characterised as having arisen from a small number of components.
%This means that the idealized conditions under which our experiments were generated will partially break down.
Nevertheless, as demonstrated by our experiment in which we used a smaller number of NMF components than the number of components, even when the data-generating process is not perfectly satisfied by observations, NMF still retains the ability to uncover meaningful components in the form of archetypical components.

A more important weakness of this work is that our experiments only considered scenarios in which observations were created from components of similar column densities and thus in which all components contributed more or less equally to the line-intensity maps. If components have dissimilar column densities then it becomes much more challenging for NMF to recover the fainter components. This is because as the lower column density components emit noticeably more weakly, correctly fitting them does not contribute much to the minimisation of the loss function. For synthetic observations containing low-column density components, in order to recover the weakest components it is crucial that the signal-to-noise within observations be high enough for these weakest components to not be drowned by noise.

%\textcolor{red}{Our experiments only considered scenarios in which gas components had similar column densities. However, we caution that the performance of NMF can be expected to degrade when components have dissimilar column densities as lower column density components, which emit more weakly, would have a much smaller contribution to the loss function. }

%In astronomical observations, this assumption of the observed emission being produced by N distinct components is typically untrue. Instead, a spectrum of gas conditions will contribute to the emission; the majority of which may be well approximated by N distinct components. Astronomers use this assumption to fit data and the NMF procedure described in this chapter presents a method of retrieving these components without bias from the astronomer. In light of this, if one NMF component's contribution to the overall emission is negligible, it makes sense to fit one fewer component to the data rather than attempting to force the algorithm to retrieve a negligible NMF component which may not reflect any real gas component. Additionally, on real observations it may be especially hard to retrieve weak components as the NMF algorithm will prioritize obtaining a good fit to the high-emission regions before attempting to fit the much weaker low column density regions.

We also wish to emphasise that our work provides only one, amongst potentially many, approaches for incorporating matrix factorization constraints into the component retrieval process. A potential extension of this work would be to combine NMF with radiative-transfer codes to enforce on retrieved components both the matrix factorization constraints of NMF and the physical realism constraints of radiative-transfer codes. 
%This could perhaps be done through a two-stage approach in which Bayesian Non-negative matrix factorization \citep{BayesianNMF} - a probabilistic extension of NMF - is first used to find a probability distributions functions over components plausibly matching observations and RADEX is secondly used to weight this distribution (which could for example be done through rejection sampling) based on a likelihood of such phases being produced according to RADEX. Application of this procedure would result in a posterior capturing both RADEX and factorization constraints. Alternatively, geometrical constraints could be incorporated into the NMF factorization, as was for example done in \cite{scarlet}.

Regardless of the exact approach taken, the large degeneracies associated with radiative transfer currently make it  difficult to constrain the properties of the ISM especially at high-redshift. Combining the information available across all spatial locations during inference, as is done in NMF and other matrix factorization approaches, is, we believe, a promising approach to address such degeneracies. However, whilst NMF is useful for understanding the relationship between lines within observations, on its own it is not sufficient for fully constraining components.

%However, we anticipate that in real-world scenarios retrieving fainter components will be even harder than suggested by such experiments. This is because for real-world observations, the idealized assumption of emission being well-approximated by a small number of components will no longer be true, and so further NMF components may be dedicated to an improved fit of the high column-density regions instead of capturing the low column-density regions as these low column-density regions will only have a minimal impact on the loss function. This means that the algorithm may not perform well when their relative contribution to the overall emission is negligible.

\section*{Acknowledgements}
The authors thank the anonymous referee for their constructive comments which significantly improved the paper. DdM and J. Heyl are funded by an STFC studentship in Data-Intensive Science (grant number ST/P006736/1). This work was  supported by the European Research Council (ERC) Advanced Grant MOPPEX 833460.

\bibliography{damien,references}{}
\bibliographystyle{aasjournal}
\section{Appendix}
\subsection{Pseudo-code}\label{appendix:pseudocode}

\iffalse
\begin{algorithm}
\caption{Algorithm for generating binned density plots}\label{alg:cap}
\begin{algorithmic}
    \For{\textbf{each} regularization \textbf{in} [W, V, W\&V]}
        \For{\textbf{each} $\alpha$ \textbf{in range}(0, 1)}
            \For{$i$ \textbf{in range}(300)}
                \State Initialize random matrix
                \State Perform NMF
                \State Compute MAE of line intensities
            \EndFor
        \EndFor
    \EndFor
\end{algorithmic}
\end{algorithm}
\fi

\begin{figure}[h]
\includegraphics[width=18cm]{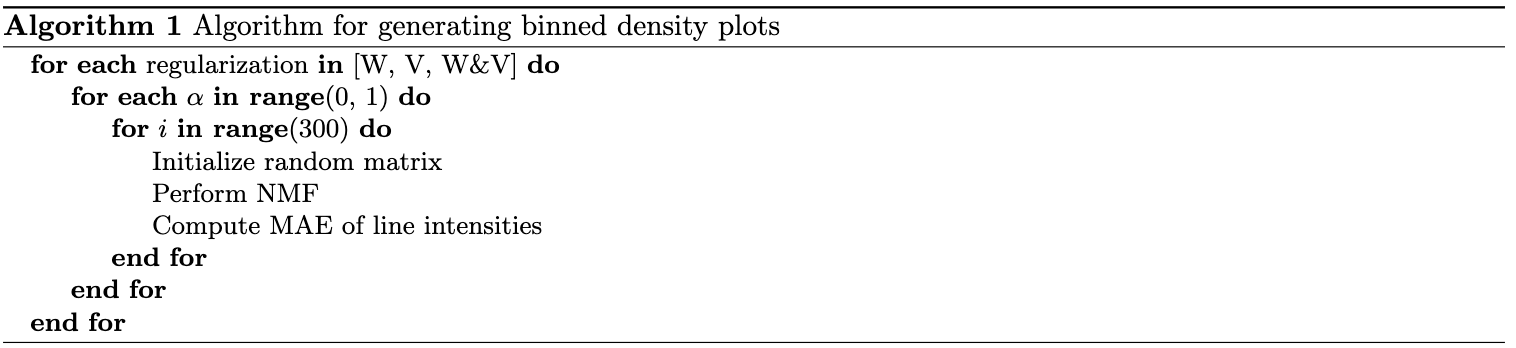}
\end{figure}



\end{document}